\documentclass[12pt]{iopart}
\usepackage{graphicx}
\usepackage{graphics}
\usepackage{subfigure}
\usepackage{multirow}
\usepackage{wrapfig}
\usepackage{float}
\usepackage{cite}
\usepackage{amssymb}
\usepackage{calrsfs}

\begin{document}
\title{Gyrobunching and wave-particle resonance in the lower hybrid drift instability}
\author{J W S Cook$^{1}$, R O Dendy$^{2,1}$ and S C Chapman$^{1}$}
\address{$^{1}$Centre for Fusion, Space and Astrophysics, Department of Physics, Warwick University, Coventry CV4 7AL, U.K.}
\address{$^{2}$Euratom/CCFE Fusion Association, Culham Science Centre, Abingdon, Oxfordshire OX14 3DB, U.K.}
\ead{j.w.s.cook@warwick.ac.uk}

\begin{abstract}
We report a first principles study of the coupled evolution of energetic ions, background majority ions, electrons and electromagnetic fields in magnetised plasma during the linear phase of the lower hybrid drift instability. A particle-in-cell code, with one spatial and three velocity space co-ordinates, is used to analyse the evolving distribution of a drifting ring-beam population of energetic protons in physical space and gyrophase angle. This analysis is carried out for plasma parameters that approximate to edge conditions in large tokamaks, in a scenario that is motivated by observations of ion cyclotron emission and may be relevant to alpha channelling. Resonant energy transfer occurs at the two gyrophase angles at which the instantaneous speed of an energetic proton on its cyclotron orbit precisely matches the phase velocity of the lower hybrid wave along the simulation domain. Electron space-charge oscillations determine the wavelength of the propagating lower hybrid wave, and thereby govern the spatial distribution of gyrobunching of the energetic protons that drive the instability.
\end{abstract}


\section{Introduction}

Understanding the nature of the wave-particle interaction that drives the lower hybrid drift instability (LHDI)\cite{ref:davidson1977physfluids,ref:hsia1979physfluids,ref:silveira2002physreve,ref:daughton2004prl} is of fundamental interest in plasma physics. The LHDI is potentially operative wherever drifting populations of ions (whether suprathermal minorities, or arising from equilibrium gradients) occur in magnetised plasmas. There is an extensive literature documenting its many roles in both space \cite{ref:dobe1999prl,ref:yoon1994physfluids,ref:drake1983physfluids} and, increasingly, laboratory \cite{ref:gladd1976plasphys,ref:carter2001prl} plasma physics. Predominantly electrostatic waves, propagating quasi-perpendicular to the magnetic field, are preferentially excited by the LHDI. Their frequency is characterised by the lower hybrid frequency $\omega_{LH}=({\Omega_{ce}\Omega_{ci}/({1+\Omega_{ce}\Omega_{ci}/\omega_{pi}^2}}))^{1/2}$, which exceeds the cyclotron frequency of the ions and is below that of the electrons. It is clear from the foregoing that the LHDI possesses features that invite deeper investigation. The instability arises from the beam-like (i.e., directed in velocity space) character of the fast ion population that drives it, while the waves excited are conditioned by the cyclotronic (i.e., rotational in velocity space) natural frequency $\Omega_{ci}$ of the background thermal ions. The waves excited are conditioned by the electron-ion mass ratio through $\Omega_{ci}/\Omega_{ce}$, implying that electron inertia plays a role. The waves excited are also conditioned by the density of charge carriers through the ion plasma frequency $\omega_{pi}$, implying that charge separation plays a central role.

In the present paper we study the physics of the excitation mechanism of the LHDI, using a particle-in-cell (PIC) code \cite{ref:devine1995jgr,ref:oppenheim1999prl,ref:dieckmann1999physplas,ref:birch2001physplas,ref:birch2001grl} which provides a fully kinetic description of the interaction between minority ($\sim1\%$) energetic protons, background thermal deuterons, electrons, the self-consistent electromagnetic fields, and the imposed background magnetic field. In particular we focus on the dynamics of the ions as they drive the LHDI, which is found to exhibit coherent bunching that is highly structured in both velocity space and real space. To anticipate some key results, the kinetic treatment of both electrons and ions enables us to capture two essential features of the physics that are not addressed (because they are averaged over) in higher order reduced models. First, the code directly represents the excited waves as propagating electron charge density oscillations with well defined frequency and wavelength. Second, the code directly represents ion motion and can therefore resolve the angular sector of an ion cyclotron orbit during which the ion is transiently in Doppler-shifted phase resonance with the wave that is being excited.

The scenario for the LHDI considered here is motivated by observations of radiative collective instability (ion cyclotron emission, ICE) of minority fusion-born and beam-injected energetic ions in the outer mid-plane edge plasma of the JET \cite{ref:cottrell1988prl,ref:cottrell1993nuclfusion,ref:mcclements1999prl} and TFTR \cite{ref:dendy1995nuclfusion,ref:cauffman1995nuclfusion,ref:mcclements1996physplasmas} tokamaks. Analysis of the spatial drift orbits of the ions responsible for exciting the fast Alfv\'{e}n waves observed showed \cite{ref:cottrell1988prl,ref:cauffman1995nuclfusion} that their velocity space distribution can be approximated by the analytical model $f(v_{\perp}, v_{\parallel}) = 1/(2\pi v_r)\delta(v_{\perp}  -  v_r) \delta (v_{\parallel} -  u)$ where $v_r$ and $u$ correspond to a pitch angle just inside the trapped-passing boundary for an ion near its birth energy. In Refs.\cite{ref:cook2010prl,ref:cook2010ppcf} it was shown that these distributions can also be unstable against the LHDI. Furthermore the lower hybrid waves excited were found to undergo Landau damping on resonant electrons asymmetrically with respect to the distribution of parallel velocities, resulting in the spontaneous creation of a current supported by suprathermal electrons\cite{ref:cook2010prl,ref:cook2010ppcf}. This is a key building block of alpha channelling \cite{ref:fisch1992prl,ref:fisch2010jplasphys} scenarios for fusion plasmas. The principle of lower hybrid current drive by Landau damping of externally applied waves was identified by Fisch and colleagues\cite{ref:fisch1978prl,ref:karney1985physreva} for fusion plasmas, and an application to spacecraft and rocket-borne measurements of electron distributions in Earth's auroral zone was noted in Ref. \cite{ref:dendy1995jgr}. The JET and TFTR fusion plasma results motivate the present choice of: plasma parameters, approximating to tokamak edge conditions, where computationally affordable; and energetic ion velocity distribution, relevant to ICE observations \cite{ref:cottrell1988prl,ref:cottrell1993nuclfusion,ref:dendy1995nuclfusion,ref:cauffman1995nuclfusion,ref:mcclements1996physplasmas,ref:mcclements1999prl} and to recent  simulations of LHDI alpha channelling\cite{ref:cook2010prl,ref:cook2010ppcf}. While these represent only one possible realisation of an LHDI scenario, among many, we believe that the underlying physics of the phase resonant excitation described in the present paper is generic to a significant extent.

\section{Fundamentals of the simulation}

Figure 1 depicts the unperturbed spatiotemporal orbit, during one cyclotron period, of a single energetic proton in a field-aligned coordinate system, and a coordinate system which is tilted with respect to the magnetic field. The initial and final projection of the equilibrium cyclotron orbit on the plane perpendicular to B is also shown for guidance. This figure shows how the energetic ion distribution $f(v_{\perp}, v_{\parallel}) \sim \delta(v_{\perp} - v_r) \delta(v_{\parallel} - u)$ governs the movement of each ion along the simulation domain $x$, and the gyrophase $\alpha = \arctan(v_{\perp,1}/v_{\perp,2})$. There are two directions of spatial variation in two co-ordinate systems shown in the figure. Only one of these is the single spatial direction $x$ of the PIC code; the other three are for schematic purposes. The $x$-projected motion is shown as a function of gyrophase $\alpha$ in Fig.2, which implies that the particle velocity $v_x = dx/dt$ along the simulation domain varies continuously, so that phase resonance between the particle $v_x$ and any wave that can propagate with phase velocity $\omega/k_x$ will only occur transiently, along a particular segment of its cyclotron orbit. It is in this segment that the perturbed $(x, \alpha)$ phase orbit of the ion resonantly exciting the wave will deviate most strongly from the unperturbed orbit of Fig.2.

\begin{figure}[H]
  \begin{center}
\includegraphics[]{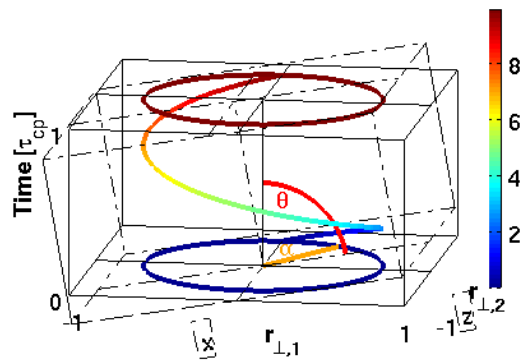}
\end{center}
\begin{small}
\caption{\label{fig:1}Unperturbed motion through time (indicated by shading (colour online)) of an energetic proton, plotted with respect to both magnetic field-based (upright solid cube) and Cartesian (tilted dashed cube) spatial co-ordinates. The plot follows one cyclotron period and incorporates drift antiparallel to {\bf{B}}, which is inclined at $\theta = 84^\circ$ to the one-dimensional spatial simulation domain $x$. Here $r_{\perp,1}$ and $r_{\perp,2}$ are orthogonal co-ordinates perpendicular to {\bf {B}}. The gyrophase $\alpha$ is zero when the proton passes through the plane defined by $x$ and {\bf {B}}. Projections of the cyclotron motion on the plane perpendicular to {\bf {B}} are shown at $t = 0$ and one cyclotron period later. The combination of cyclotron motion and drift parallel to {\bf {B}} results in the drift shown in Fig.2.}
\end{small}
\end{figure}

\begin{figure}[H]
\begin{center}
\includegraphics[]{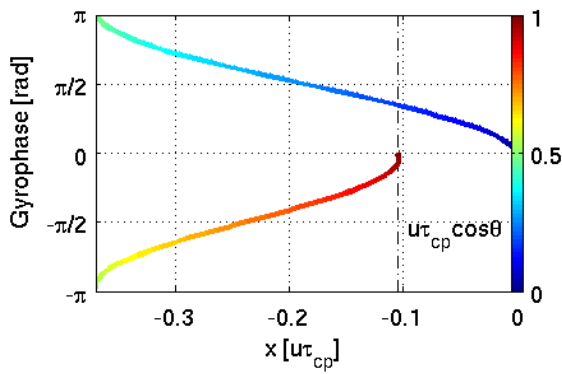}
\end{center}
\begin{small}
\caption{\label{fig:2}Projected motion through time (indicated by colour) of an energetic proton along the simulation domain direction $x$ as a function of gyrophase $\alpha$. This motion is implicit in the definition of the unperturbed velocity distribution in terms of $v_{\perp}$ and $v_{\parallel}$, and follows from Fig.1. The parallel velocity component present in this geometry causes a drift in $x$ of $u\tau_{cp}\cos\theta$ per gyro-orbit, where $\tau_{cp} = 2\pi/\Omega_{cp}$ is the proton gyroperiod.}
\end{small}
\end{figure}

It is known from Figs.3 and 4 of Ref.\cite{ref:cook2010prl} that the dominant collectively excited modes propagate in the negative $x$-direction with $\omega/k_x  \sim 1.5\times10^{-7} m/s \lesssim  v_x = u\cos\theta - v_r\sin\theta$. Modes with slightly lower amplitude are excited propagating in the opposite direction with similar phase speed. These phase velocities are located near the maxima of transient speeds, in either direction along the simulation domain, of the energetic ions that drive the instability. These occur (see Figs. 1 and 2) at gyrophase $\alpha \sim \
\pi/2$ and $\sim 3\pi/2$; see also panel (a) of Fig.2 of Ref.\cite{ref:cook2010ppcf}.

\section{Simulation results}

We now turn to the resonant wave-particle interaction as manifested by the distribution of energetic protons along the physical co-ordinate $x$ and with respect to gyrophase angle $\alpha$. Figure 3 plots (left) the whereabouts of particles in $(x,\alpha)$ phase space at $t = 19.8\tau_{LH}$, towards the end of the linear phase of the LHDI in this scenario, and (right) the corresponding distribution of particle energies. Within the spatial limits of the simulation box, there are 21 gyrophase bunched structures which span the whole gyrophase range: 10.5 of these are visible in Fig 3, because data from only half of the simulation domain is shown (Figs 3 to 6 all show data from only half of the spatial domain for reasons of pictorial resolution). The simulation domain length $L \simeq 6.3$ proton Larmor radii. This is sufficiently large that protons undergo $\sim7$ gyro-orbits per boundary crossing, which corresponds to approximately one boundary crossing per proton during the entire simulation. Figure 3 demonstrates that the rapidly varying behaviour of the distribution in the negative $\alpha$ domain of phase space is a continuation of the slower variation in the positive domain. This new result is not apparent from previous analysis of the LHDI in this regime, for example Fig.5 of Ref.\cite{ref:cook2010prl} and Fig.4 of Ref.\cite{ref:cook2010ppcf}. 

\begin{figure}[H]
  \begin{center}
    \subfigure[]{\label{fig:3a}\includegraphics[]{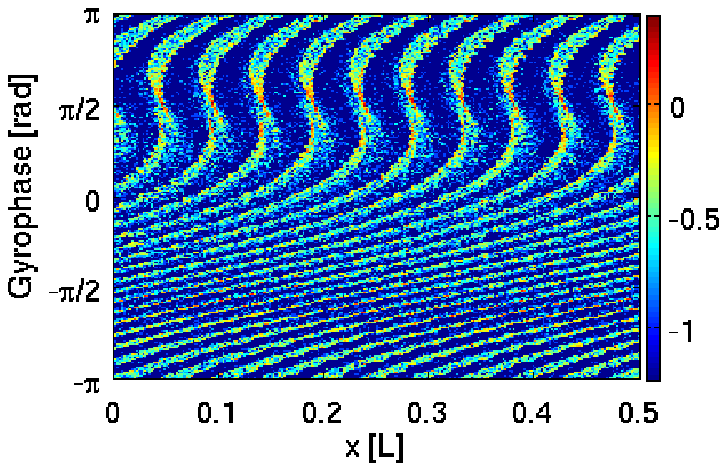}}
    \subfigure[]{\label{fig:3b}\includegraphics[]{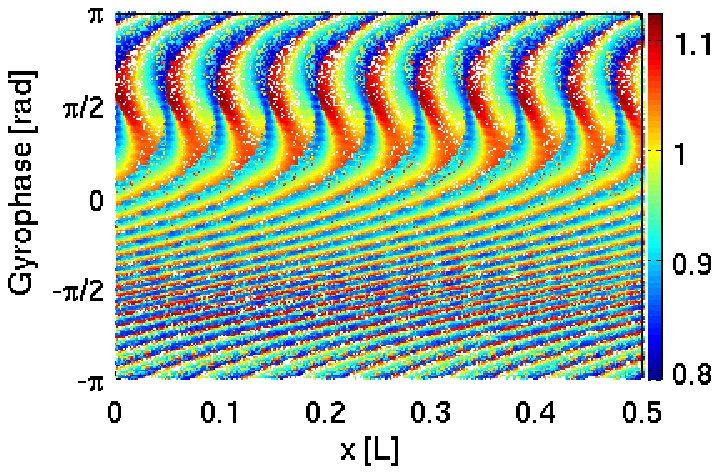}}
  \end{center}
\begin{small}
\caption{\label{fig:3}PIC code derived proton population data from a snapshot at $t = 19.8\tau_{LH}$ towards the end of the linear stage of the instability, plotted as a function of gyrophase (ordinate axis) and position (abscissa). Position axes are in units of box length $L$ and gyrophase axes are in radians. Panel (a): Proton probability density on a $log_{10}$ scale indicated by shading (colour online). Panel (b): Proton energy in units of $\varepsilon_{p}(t=0)$ indicated by shading (colour online). Both panels show 10.5 nearly-identical {\it{s}}-shaped features in the upper half plane at regular intervals in the spatial direction, as well as smaller scale repeating structures.}
\end{small}
\end{figure}

There are 10.5 gyrophase bunched structures per half-box because the simulation box accommodates 21 wavelengths of the wave that is found to be spontaneously excited. This is demonstrated in Fig. 4, which plots the spatial distribution of electrons at the snapshot in time shown in Fig.3, and also confirms that the predominantly electrostatic excited wave is  primarily a charge density oscillation supported by the electrons.

\begin{figure}[H]
\begin{center}
\includegraphics[]{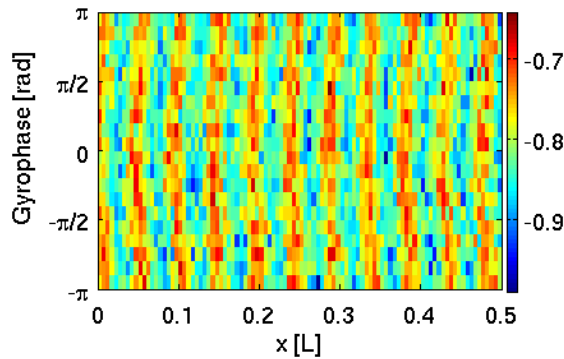}
\end{center}
\begin{small}
\caption{\label{fig:4}PIC code derived electron population data from a snapshot at $t = 19.8\tau_{LH}$ towards the end of the linear stage of the instability. Electron probability density on a $log_{10}$ scale is indicated by shading (colour online) as a function of gyrophase (ordinate axis) and position (abscissa). Position axes are in units of box length $L$ and gyrophase axes are in radians. The vertical stripes indicate that electron bunching occurs as a function of position and not gyrophase: electrons participate in wave action by charge density oscillation.}
\end{small}
\end{figure}

Figure 3 (left) captures wave excitation by the energetic ions as it occurs. The strong spatially dispersive distortion at $\alpha = \pi/2$ corresponds (see Fig.1) to the maximum negative particle velocity (see also Fig.2(a) of Ref.\cite{ref:cook2010ppcf}), which is located between the points in gyrophase where the resonant drive of the dominant backward-propagating waves takes place. The speed of the dominant wave is slightly smaller than the maximum speed of the protons. Consequently there are two positions of resonance in gyrophase, either side of the maximum negative velocity.  These resonant perturbations apparently condition the long-wavelength (in gyrophase $\alpha$) structure across the positive-$\alpha$ domain of phase space. At $\alpha = -\pi/2$, which corresponds to the maximum positive particle velocity, the consequences of resonance are apparent in the moir{\'e} pattern (which is more easily visible in Fig.3 (right)). While both these velocity resonances give rise to significant (but differing by an order of magnitude) concentrations of electromagnetic field energy in ($\omega,k$)-space, as shown in Fig.3 of Ref.\cite{ref:cook2010prl}, it follows from Fig.3 that the phase space consequences are different in the two cases. The number of moir{\'e} s-shaped features in the lower half plane of Fig.3 exceeds that in the upper half plane. This reflects the difference in $k_x$ for the waves excited at the two resonances, see again Fig. 3 of Ref.\cite{ref:cook2010prl}.

Figure 3 (right) assists interpretation of the physics in terms of the simplified analytical model for the dynamics of the particle-wave interaction that was introduced in Eqs.(12) to (16) of Ref.\cite{ref:cook2010ppcf}. The essence of this model is that the impact of wave-particle interaction, for a proton with initial gyrophase $\alpha_0$ at $t = 0$ and at arbitrary $(x,t)$, scales with the magnitude of the electric field which that proton experienced when it was most closely in phase resonance with the wave. This resonance will have occurred at time $t_R(\alpha_0)$ such that $v_x(t_R) = \omega/k$, where $\omega \sim 6\omega_{LH}$ and $k = - 2\omega_{pe}/c$ are the angular frequency and wavenumber of the dominant excited mode in the simulation, obtained from the analysis of the Fourier transform of the electromagnetic field in Figs.3 of Refs.\cite{ref:cook2010prl}. Given an energetic proton moving on an unperturbed orbit drawn from the drifting beam distribution specified previously, expressions for $t_R$ and position $x_R(\alpha_0,x,t)$ follow from Eqs.(14) and (15) of Ref.\cite{ref:cook2010ppcf}. The key point is that the field amplitude, $\mathcal{E}(\alpha_0,x,t)$ experienced at the point of resonance $(x_R,t_R)$, which is given by Eq.(16) of Ref.\cite{ref:cook2010ppcf}, is considered a property of the proton at all subsequent $(x,t)$. Figure 5 (left) plots $\mathcal{E}(\alpha_0,x,t)$ for a population of energetic protons that is initially uniformly distributed in space $x$ and gyrophase angle $\alpha_0$ at an arbitrary time $t = 2\pi/\Omega_{cp}$. Figure 5 (right) is derived from the analytical model using the three waves identified from the Fourier transform of the excited electromagnetic field that is generated by the PIC simulation during its linear phase. To construct it we repeat the operation used for Fig. 5 (left) while adding: a second wave whose $(\omega,k)$ are those identified as the harmonic of the dominant backward travelling wave in the Fourier transform $(\omega_h = 12\omega_{LH}, k_h = -4 \omega_{pe}/c)$, and whose amplitude is smaller by a factor ten; and a third wave whose $(\omega,k)$ are those identified for the forward travelling excited wave in the Fourier transform $(\omega_f = 6.6\omega_{LH}, k_f = 1.3\omega_{pe}/c)$, and whose amplitude is smaller by a factor four. 

Figure 5 (right) appears to capture the key features of Fig.3 (right). This analytical model is based on unperturbed proton gyro-orbits, and Fig. 5 shows that these suffice to recreate gyrobunching effects seen in the PIC simulations provided perturbations are small in the sense that: the resonant gyrophase angles are sharply defined; and the spatial deviation of the resonant particles from their unperturbed orbits is small compared to the wavelength.

This analytical model is based on unperturbed proton gyro-orbits and is able to recreate gyrobunching effects seen in the PIC simulation. This is the case because perturbations are small, in the sense that the resonant gyrophase angles are sharply defined and the spatial deviation (as distinct from gyrophase deviation) of the resonant particles from their unperturbed orbits is also small.

\begin{figure}[H]
  \begin{center}
    \subfigure[]{\label{fig:5a}\includegraphics[]{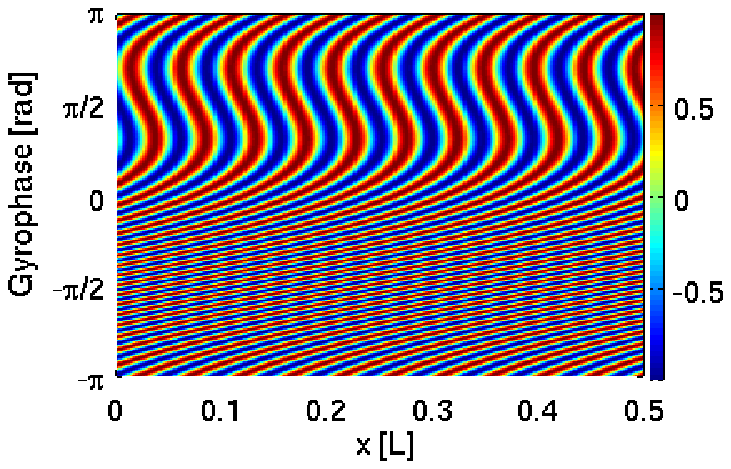}}
    \subfigure[]{\label{fig:5b}\includegraphics[]{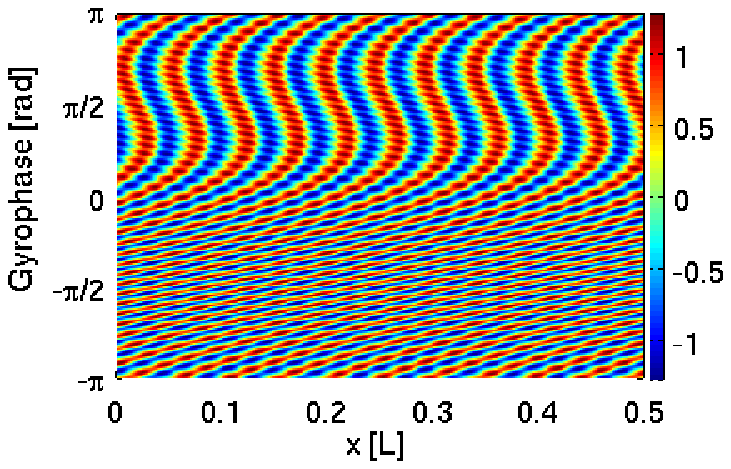}}
  \end{center}
\begin{small}
\caption{\label{fig:5}Normalized electric field amplitude $\mathcal{E}$ indicated by shading (colour online) experienced by protons at their most recent point of resonance, as a function of proton gyrophase and position at a snapshot in time (see text). Panel (a): Normalized wave amplitude seen by protons at resonance with the dominant wave. Panel (b): As panel (a) for the sum of dominant wave of unit amplitude, its second harmonic of one tenth amplitude, and a wave travelling in the opposite direction of one quarter amplitude. Data in panel (a) shows the effect of resonance with the dominant  exited wave as visible from the large scale structure in Fig. 3a and b, while panel (b) shows finer scale structures, as in Fig. 3b, highlighting the effects of resonances with less powerful excited waves.}
\end{small}
\end{figure}

\section{Conclusions}

Analysis of the distribution in physical space and gyrophase angle of a drifting ring-beam population of energetic ions, as they drive the linear phase of the lower hybrid drift instability in a 1D3V PIC simulation, yields insights into the underlying plasma physics. The spatial coherence and spatial variation of the gyrophase bunched structures are shown to reflect a combination of cyclotron-type and beam-type dynamical and resonant effects, conditioned by the interplay of ion and electron dynamics. These new results shed fresh light on the hybrid character of the lower hybrid drift instability.

Resonant energy transfer occurs at the narrowly defined gyrophases at which the instantaneous speed of an energetic proton, from the drifting ring-beam population, on its cyclotron orbit precisely matches the phase velocity of the lower hybrid wave along the simulation domain. As in any resonant wave-particle instability, coupling to a wave that can propagate is necessary. In the present case, it is electron space-charge oscillation (Fig. 4) that determines the wavelength of the propagating lower hybrid wave, and thereby governs the spatial distribution of gyrobunching of the energetic protons (Figs. 3 and 5) that drive the instability. The gyrophase structure of phase space bunching at any given position is governed by the proton trajectories on their cyclotron orbits, which are oblique to the direction of propagation of waves along the simulation domain,  in conjunction with restrictions on the spatial separation of gyrobunches due to collective electron oscillation.

Many of the key features of the wave-particle interaction that we observe in the simulations can be understood retrospectively in terms of a simplified analytical model. However this model is wholly reliant on PIC code outputs for its specification, including the number of waves, their relative amplitudes, and their frequencies and wavenumbers.

This analysis was carried out for plasma parameters that approximate, where computationally affordable, edge plasma conditions in a large tokamak. While the physics is generic, we note that this scenario is motivated by observations of ion cyclotron emission from JET\cite{ref:cottrell1988prl} and TFTR  \cite{ref:cauffman1995nuclfusion}, and may be relevant to alpha channelling\cite{ref:cook2010prl,ref:cook2010ppcf}.

\ack
This work, part-funded by the European Communities under the contract of Association between EURATOM and CCFE, was partly carried out within the framework of the European Fusion Development Agreement. The views and opinions expressed herein do not necessarily reflect those of the European Commission. This work was also part-funded by the RCUK Energy Programme under grant EP/I501045. This project used the EPOCH code developed under EPSRC grant EP/G054950/1 and the authors thank Christopher Brady and the EPOCH development team. The authors also thank Nathaniel Fisch for helpful conversations.

\section*{References}

\begin{small}
\bibliographystyle{unsrt}
\bibliography{bibliography}
\end{small}

\end{document}